# A SHARPER DISCREPANCY MEASURE FOR POST-ELECTION AUDITS

BY PHILIP B. STARK

*University of California—Berkeley*

Post-election audits use the discrepancy between machine counts and a hand tally of votes in a random sample of precincts to infer whether error affected the electoral outcome. The maximum relative overstatement of pairwise margins (MRO) quantifies that discrepancy. The electoral outcome a full hand tally shows must agree with the apparent outcome if the MRO is less than 1. This condition is sharper than previous ones when there are more than two candidates or when voters may vote for more than one candidate. For the 2006 U.S. Senate race in Minnesota, a test using MRO gives a $P$-value of 4.05% for the hypothesis that a full hand tally would find a different winner, less than half the value Stark [*Ann. Appl. Statist.* **2** (2008) 550–581] finds.

**1. Maximum relative overstatement of pairwise margins.** For a candidate other than an apparent winner to be a real winner of an election, error that hurts that candidate or helps an apparent winner must exceed that apparent winner's margin of victory over *that* candidate. The maximum relative overstatement of pairwise margins (MRO) takes that into account; previous measures compare errors with the margin of victory over the runner-up alone.

Consider a contest with $K$ candidates, $1, \ldots, K$, and $N$ precincts, $1, \ldots, N$. Each voter may vote for up to $f$ candidates. The $f$ candidates who apparently won are those in $\mathcal{K}_w$. Those who apparently lost are in $\mathcal{K}_\ell$. The apparent vote for candidate $k$ in precinct $p$ is $v_{kp}$. The apparent vote for candidate $k$ is $V_k \equiv \sum_{p=1}^N v_{kp}$. The apparent margin of candidate $w$ over candidate $\ell$ is $V_{w\ell} \equiv V_w - V_\ell$. For $w \in \mathcal{K}_w$ and $\ell \in \mathcal{K}_\ell$, $V_{w\ell} > 0$: the apparent winners are the $f$ candidates with strictly positive apparent margins over the other $K - f$.

*Actual*, as a modifier of vote, margin, winner or electoral outcome, means what a full hand tally would show. The actual vote for candidate $k$ in precinct $p$ is $a_{kp}$. The actual vote for candidate $k$ is $A_k \equiv \sum_{p=1}^{N} a_{kp}$. The actual margin of candidate $w$ over candidate $\ell$ is $A_{w\ell} \equiv A_w - A_\ell$. The apparent winners are the actual winners if

$$\min_{w \in \mathcal{K}_w, \ell \in \mathcal{K}_\ell} A_{w\ell} > 0. \tag{1}$$

Define

$$e_{pw\ell} \equiv \frac{(v_{wp} - v_{\ell p}) - (a_{wp} - a_{\ell p})}{V_{w\ell}}. \tag{2}$$

For the apparent and actual electoral outcomes to differ, there must exist $w \in \mathcal{K}_w$ and $\ell \in \mathcal{K}_\ell$ for which $\sum_{p=1}^{N} e_{pw\ell} \geq 1$. The *maximum relative overstatement of pairwise margins (MRO) in precinct $p$* is

$$e_p \equiv \max_{w \in \mathcal{K}_w, \ell \in \mathcal{K}_\ell} e_{pw\ell}. \tag{3}$$

Now

$$\max_{w \in \mathcal{K}_w, \ell \in \mathcal{K}_\ell} \sum_{p=1}^{N} e_{pw\ell} \leq \sum_{p=1}^{N} \max_{w \in \mathcal{K}_w, \ell \in \mathcal{K}_\ell} e_{pw\ell} = \sum_{p=1}^{N} e_p. \tag{4}$$

The sum on the right is the *maximum relative overstatement of pairwise margins (MRO)*. If the apparent and actual electoral outcomes differ, $\sum_{p=1}^{N} e_p \geq 1$. When $K = 2$ and $f = 1$, this is equivalent to the condition Stark (2008) tests. But for $K > 2$ or $f > 1$, this condition can be much sharper.

Suppose the number of valid ballots cast in precinct $p$ is at most $b_p$. Clearly, $a_{wp} \geq 0$ and $a_{\ell p} \leq b_p$. Hence, $e_{pw\ell} \leq (v_{wp} - v_{\ell p} + b_p)/V_{w\ell}$, and so

$$e_p \leq \max_{w \in \mathcal{K}_w, \ell \in \mathcal{K}_\ell} \frac{v_{wp} - v_{\ell p} + b_p}{V_{w\ell}} \equiv u_p. \tag{5}$$

Let $\{w_p(\cdot)\}_{p=1}^{N}$ be monotonic functions. Stark's (2008) method can test the hypothesis that $\sum_{p=1}^{N} e_p \geq 1$ given the constraint $e_p \leq u_p$ using the maximum observed value of $w_p(e_p)$ as the test statistic: substitute $M = 1$ and the definitions of $u$ and $e_p$ given here.

**2. The 2006 U.S. Senate race in Minnesota.** Table 1 lists the vote totals for the 2006 U.S. Senate race in Minnesota. The apparent winner was Amy Klobuchar.[1] Minnesota elections law S.F. 2743 (2006) requires that counties with fewer than 50,000 registered voters audit at least two precincts chosen at random; that counties with between 50,000 and 100,000 registered voters audit at least three; and that counties with more than 100,000 registered voters audit at least four. At least one precinct audited in each county must have 150 or more votes. Minnesota has 4,123 precincts in 87 counties. 202 precincts were audited after the 2006 election. Several counties audited more than the legal minimum.

---

[1] See www.sos.state.mn.us/docs/2006_General_Results.XLS, electionresults.sos.state.mn.us/2006 1107/ElecRslts.asp?M=S&Races=0102, and www.sos.state.mn.us/home/index.asp?page=544.

Table 1
*Summary of 2006 U.S. Senate race in Minnesota*

| Voters | Undervotes & invalid ballots | Fitzgerald (Indep) | Kennedy (Repub) | Klobuchar (Democ/Farm/ Labor) | Cavlan (Green) | Powers (Constit) | Write-ins |
|---|---|---|---|---|---|---|---|
| 2,217,818 | 15,099 | 71,194 | 835,653 | 1,278,849 | 10,714 | 5,408 | 901 |
| $V_{w\ell}$ | N/A | 1,207,655 | 443,196 | N/A | 1,268,135 | 1,273,441 | 1,277,948 |

Following Stark (2008), we pool Cavlan, Powers, and the write-in candidates to form a pseudo-candidate who apparently lost to Klobuchar by 1,261,773 votes. Thus, $K = 4$, $f = 1$ and $N = 4{,}123$. The maximum value of $u_p$ is 0.0097. The maximum observed value of $e_p$ is $4.5 \times 10^{-6}$. If Klobuchar actually lost, the MRO in at least 166 precincts must be larger than any in the sample. In contrast, for the measure of margin overstatement Stark (2008) uses, only about 130 precincts would need to have values exceeding any in the sample.[2] Thus, it is easier to confirm that the apparent and actual outcomes agree using the MRO.

We calculate a conservative $P$-value for the hypothesis that Klobuchar actually lost by pretending that the sample was drawn with replacement from all 4,123 precincts, but that only 78 precincts were drawn, as if the population were sampled using the minimum sampling fraction among counties.[3] For weight functions $w_p(x) = x$, the $P$-value is the maximum chance that 78 precincts chosen at random with replacement would have $e_p \leq 0.0097$ if, among all 4,123 precincts, there were at least 166 with $e_p > 0.0097$. That value is $(\frac{4123-166}{4123})^{78} = 4.05\%$, roughly half the conservative $P$-value of 8.22% Stark (2008) finds.

If 202 precincts were drawn as a simple random sample and the same discrepancies were observed, the $P$-value would be about 0.02% using the MRO. In contrast, Stark (2008) finds a corresponding $P$-value of about 0.13%.

**3. Conclusion.** The MRO yields a sharper necessary condition for the apparent electoral outcome to differ from the outcome a full hand tally would show than previous measures of the discrepancy between machine and hand counts do. An a priori bound for the MRO in a precinct can be derived from a bound on the number of valid ballots in that precinct. The testing framework Stark (2008) develops works with MRO if the definitions of $M$, $u$ and $e$ are revised, and yields a more powerful test.

---

[2]See Table 5 of Stark (2008).
[3]See Section 4.2.1 of Stark (2008).

**Acknowledgments.** I thank Kathy Dopp, Mark Lindeman and Luke Miratrix for helpful conversations.

## REFERENCE

Stark, P. B. (2008). Conservative statistical post-election audits. *Ann. Appl. Statist.* **2** 550–581.

Department of Statistics, Code 3860
University of California—Berkeley
Berkeley, California 94720-3860
USA
E-mail: stark@stat.berkeley.edu